\documentclass[preprint,showpacs,preprintnumbers,amsmath,amssymb]{revtex4}

\usepackage{graphicx}
\usepackage{dcolumn}
\usepackage{bm}

\begin{document}

\title{Pressure effect on magnetism in CeTe$_{1.82}$}

\author{M. H. Jung}
\thanks{Author to whom correspondence should be addressed.
Present address: MST, Korea Basic Science Institute, Daejeon 305-333,
Korea, E-mail : mhjung@kbsi.re.kr} \affiliation{NHMFL, Los Alamos
National Laboratory, Los Alamos NM 87545, USA}
\author{A. Alsmadi}
\affiliation{Department of Physics, New Mexico State University,
Las Cruces NM 88003, USA}
\author{H. C. Kim}
\affiliation{MST, Korea Basic Science Institute, Daejeon 305-333,
South Korea}
\author{J. Kamarad}
\affiliation{Institute of Physics, Czech Academy of Sciences, 162
53 Prague 6, Czech Republic}
\author{T. Takabatake}
\affiliation{Department of Quantum Matter, ADSM, Hiroshima
University, Higashi-Hiroshima 739-8526, Japan}

\date{\today}

\begin{abstract}
We report the normal-state transport and magnetic properties of a
pressure-induced superconductor CeTe$_{1.82}$. We found that the
applied pressure is required to increase the Kondo temperature
scale ($T^*_{\rm K} \sim$ 170 K), associated with the
two-dimensional motion of the carriers confined within the Te
plane. Both the short-range ferromagnetic ordering temperature
($T_{\rm SRF} \sim$ 6 K) and the long-range antiferromagnetic
transition temperature ($T_{\rm N} \sim$ 4.3 K) are slightly
increased with pressure. We suggest that the application of
pressure enhances a coupling between the 4$f$ and conduction
electrons. We also found that the field effect on the transport
under pressure is analogous to that at ambient pressure, where a
large magnetoresistance is observed in the vicinity of $T_{\rm
SRF}$.
\end{abstract}



\maketitle

The generally accepted picture in $f$-electron systems is that the
$f$ electrons are responsible for the local magnetic moments and
are strongly correlated with itinerant conduction electrons
\cite{HF}. Then, there are two magnetic interactions that we can
expect from the $f$-electron systems. One is the on-site Kondo
interaction, a screening of the local magnetic moment by
conduction electrons, which develops nonmagnetic ground state. The
other is the indirect RKKY exchange interaction between the
localized $f$ magnetic moments mediated by the conduction
electrons, favoring long-range magnetic ordering. Consequently,
the balance between these two competing interactions leads to
various ground states \cite{Doniach}.

A striking feature in the $f$-electron systems is that for certain
rare-earth compounds the magnetic order on the rare-earth
sublattice has been found to coexist with superconducting state
\cite{HF_SC}. Recently, the coexistence of magnetism with
superconductivity (SC) has been observed in CeTe$_{1.82}$
\cite{CeTe2_SC}, where the pairing mechanism is suggested to be a
phonon-mediated SC, associated with the charge density wave and
enhanced by the ferromagnetic fluctuations. In this paper, we
report the normal-state properties as functions of pressure and
magnetic field for CeTe$_{1.82}$. At ambient pressure, this
compound displays two different magnetic orderings. The local
magnetic moments of Ce ions develop a short-range ferromagnetic
ordering in the CeTe layer with a magnetoelastic origin below
$T_{\rm SRF} \sim$ 6 K. The short-range ferromagnetic CeTe layers
change a long-range ferromagnetic order in the layers and
simultaneously a long-range antiferromagnetic order in the spin
sequence of down-up-up-down along the $c$ axis below $T_{\rm N}
\sim$ 4.3 K \cite{Jung_phase}. Finally, there appears to be a
superconductivity below $T_{\rm c} \sim$ 2.7 K at pressure as
small as 2 kbar \cite{CeTe2_SC}.

Figure 1 shows the results of electrical resistivity $\rho(T)$ for
CeTe$_{1.82}$ with different current directions applied in the
basal plane of the tetragonal unit cell ($I \perp c$), along the
direction of 45 degrees in the plane ($I$ $\angle$ $c$), and along
the $c$ axis ($I \parallel c$). For clarity, the resistivity
values have been scaled by a factor of 4 for $I$ $\angle$ $c$ and
10 for $I \parallel c$. The absolute values are in the order of
$\rho_{\perp c} < \rho_{\angle c} < \rho_{\parallel c}$. There is
a metallic behavior for $\rho_{\perp c}$ and a semiconducting
behavior for $\rho_{\parallel c}$, reflecting the anisotropic
structure. A broad maximum at $T^*_{\rm K} \sim$ 170 K in
$\rho_{\perp c}$ is associated with the onset of the Kondo-lattice
ground states, as observed in other Ce-based intermetallics
\cite{HF}. This maximum moves to lower temperature 100 K for
$\rho_{\angle c}$ and finally disappears for $\rho_{\parallel c}$.
This result is well understood in terms of the reduced Kondo
effect by the semiconducting CeTe layers stacked along the $c$
axis. The $c$-axis semiconducting state cannot be mapped by the
Kondo picture because the Kondo effect requires a moderate carrier
density. On the other hand, a sharp peak at $T_{\rm SRF} \sim$ 6 K
is suggested as a result of short-range ferromagnetic order by
magnetic polarons \cite{Jung_phase}. Since the magnetic polarons
are coupled with carrier localization and spin polarization, it is
natural that this effect is most dominant at temperatures just
above $T_{\rm N}$ and is independent of the current direction. If
these are indeed the cases, the two features should be drastically
affected by external parameters such as applied pressure and
magnetic field.

The pressure variations of $\rho_{\perp c}(T)$ are illustrated in
Fig. 2 for the whole temperature region and in the inset for the
low-temperature region. There are three remarkable pressure
effects in CeTe$_{1.82}$. First, the resistivity curves are
crossing at temperature between $T^*_{\rm K}$ and $T_{\rm SRF}$.
An applied pressure increases the high-temperature resistivity
values but decreases the low-temperature resistivity values. This
result gives a possibility of different crystalline-electric-field
effect under pressure. Second, the applied pressure changes the
position of high-temperature Kondo feature at $T^*_{\rm K}$ to
higher temperature. This result is well explicable by considering
the enhanced Kondo interaction in an applied pressure. Third, the
low-temperature magnetic feature at $T_{\rm SRF}$ moves slowly to
higher temperature with increasing pressure. This result gives an
evidence for the idea that the number of magnetic polaron states
is increased with shrinking the crystal lattice.

The existence of magnetic polaron state can be investigated by
measuring the magnetic field dependence of $\rho_{\perp c}(T)$
under pressure. In Fig. 3, the sharp peak at $T_{\rm SRF}$ is
strongly suppressed and shifts to higher temperatures with
increasing magnetic field, as was found in the ambient-pressure
measurements \cite{Jung_phase}. This leads to a large negative
magnetoresistance (MR) in the vicinity of $T_{\rm SRF}$. As shown
in the inset of Fig. 3, MR measured at temperatures below (1.9 K)
and above $T_{\rm SRF}$ (13 K) increases monotonically, while MR
at around $T_{\rm SRF}$ (6.2 K) initially decreases with
increasing fields and then turns to increase making a broad
minimum at 3 T. This negative MR behavior at low fields is a
characteristic of the low-carrier-density materials showing
magnetic polaron states.

In order to look at the influence of pressure on the
antiferromagnetic order at $T_{\rm N}$, we have measured the
magnetization $M(T)$ in a field of 50 Oe with varying pressure,
which is shown in Fig. 4. There is a small difference between
zero-field-cooled (ZFC) and field-cooled (FC) data at around
$T_{\rm SRF}$, due to the short-range ferromagnetic ordering. A
sharp peak associated with the long-range antiferromagnetic order
at $T_{\rm N}$ increases almost linearly with increasing pressure
at a rather large rate of $dT_{\rm N}/dP$ = 0.06 K/kbar. This
implies a strong hybridization between the localized 4$f$
electrons and the conduction electrons due to the decrease of
interatomic distance by an applied pressure. This strong magnetic
interaction can cause the itinerant conduction electrons to be
superconducting, as reported previously \cite{CeTe2_SC}.

Special attention can be given to a quantum critical point (QCP)
in CeTe$_{1.82}$ by applying pressure. As expected from Doniach
phase diagram \cite{Doniach}, QCP is normally connected to nearly
heavy-fermion metals on the border of antiferromagnetic and
ferromagnetic states (N\'{e}el or Curie temperature, $T_{\rm N}$
or $T_{\rm C} \rightarrow$ 0 K) \cite{QCP} and/or at the phase
boundary of deviations from Fermi-liquid theory (non-Fermi liquid,
NFL) \cite{NFL}. In CeTe$_{1.82}$, the superconducting transition
is observed at $T_{\rm c}$ = 2.7 K well below the magnetic
ordering temperatures ($T_{\rm SRF}$ and $T_{\rm N}$), and the
normal state properties are far from the NFL behavior. Combining
these observations and no SC in other single crystals of
CeTe$_{1.85}$ and CeTe$_{1.87}$, we suggested that this
pressure-induced SC is mediated by phonon, possibly enhanced by
the ferromagnetic fluctuations inside the magnetic ordering phase
\cite{CeTe2_SC}.

In conclusion, we confirmed the crossover from the metallic
behavior in the $ab$ plane to the semiconducting behavior along
the $c$ axis. CeTe$_{1.82}$ displays various collective ground
states; Kondo-lattice coherence ($T^*_{\rm K} \sim$ 170 K) in the
metallic $ab$ plane, short-range ferromagnetic ordering (at
$T_{\rm SRF} \sim$ 6 K) in the plane, long-range antiferromagnetic
transition ($T_{\rm N} \sim$ 4.3 K) along the $c$ axis, and
finally superconducting transition ($T_{\rm c} \sim$ 2.7 K) under
pressure. We found that the Kondo interaction $T^*_{\rm K}$ is
enhanced by an applied pressure. Both $T_{\rm SRF}$ and $T_{\rm
N}$ slightly increase with pressure but decrease with magnetic
field. We thus conclude that the pressure is required to enhance
the magnetic exchange interaction and finally to make $T_{\rm c}$.

The authors thank A. H. Lacerda, H. Nakotte, H. C. Ri, and K. Umeo
for their helps in the resistivity and magnetization measurements
under pressure. This work at NHMFL was performed under the
auspices of the National Science Foundation, the State of Florida
and the US Department of Energy.

\begin{figure}
\begin{center}
\includegraphics[width=1\linewidth]{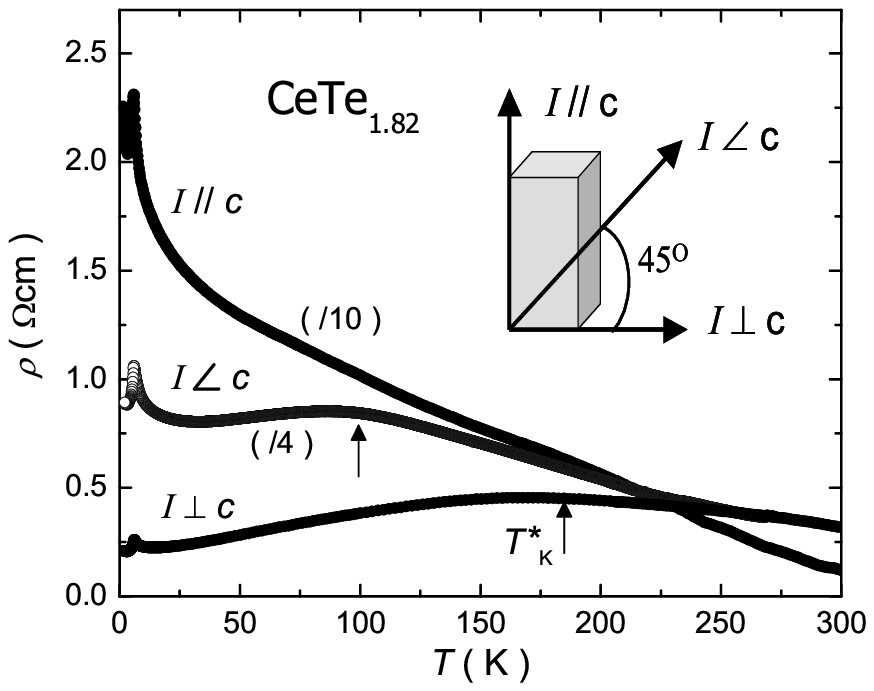}
\caption{Electrical resistivity $\rho(T)$ with different current
directions applied in the $ab$ plane ($I \perp c$), along the
direction of 45 degrees in the plane ($I \angle c$), and along the
$c$ axis ($I \parallel c$). For clarity, the resistivity values
have been scaled by a factor of 4 for $I \angle c$ and 10 for $I
\parallel c$. The arrows indicate the coherent Kondo temperature, $T^*_{\rm K}$.}
\end{center}
\end{figure}

\begin{figure}
\begin{center}
\includegraphics[width=1\linewidth]{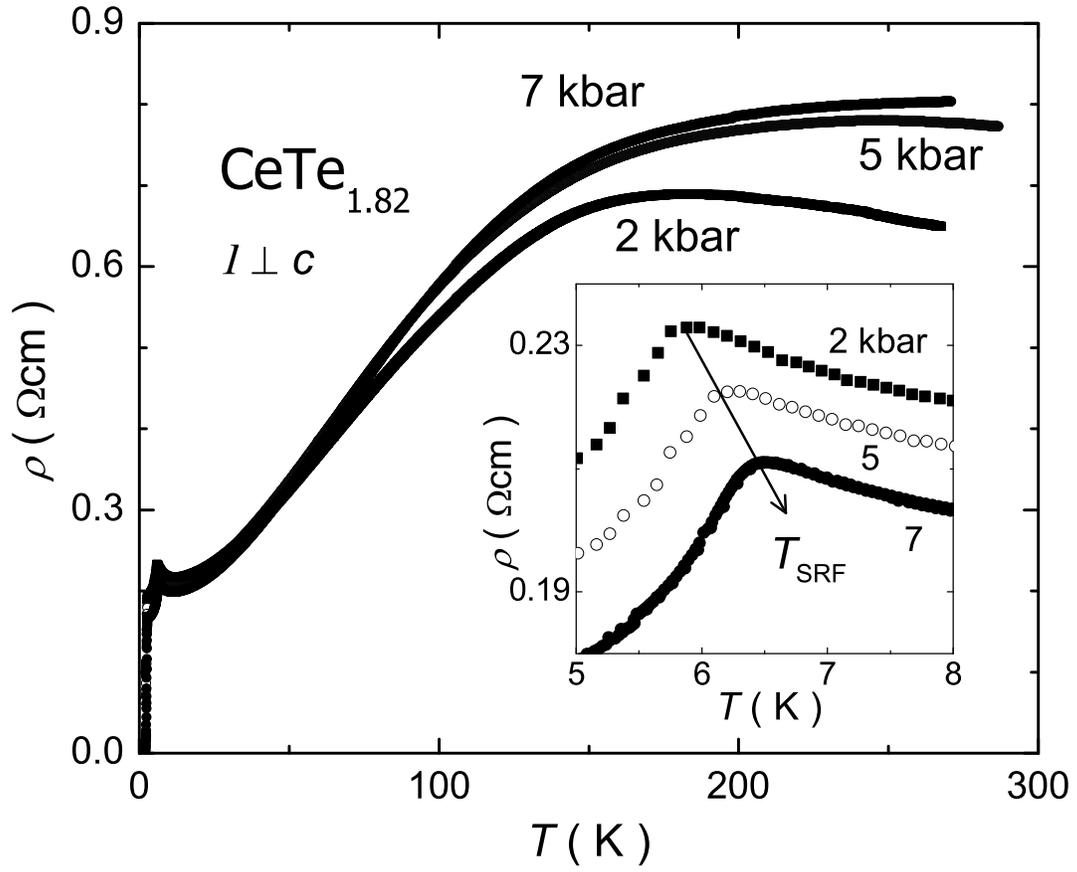}
\caption{In-plane resistivity $\rho_{\perp c}(T)$ measured at
several pressures, 2, 5, and 7 kbar. The inset shows the
low-temperature data, clarifying the short-range ferromagnetic
ordering temperature, $T_{\rm SRF}$.}
\end{center}
\end{figure}

\begin{figure}
\begin{center}
\includegraphics[width=1\linewidth]{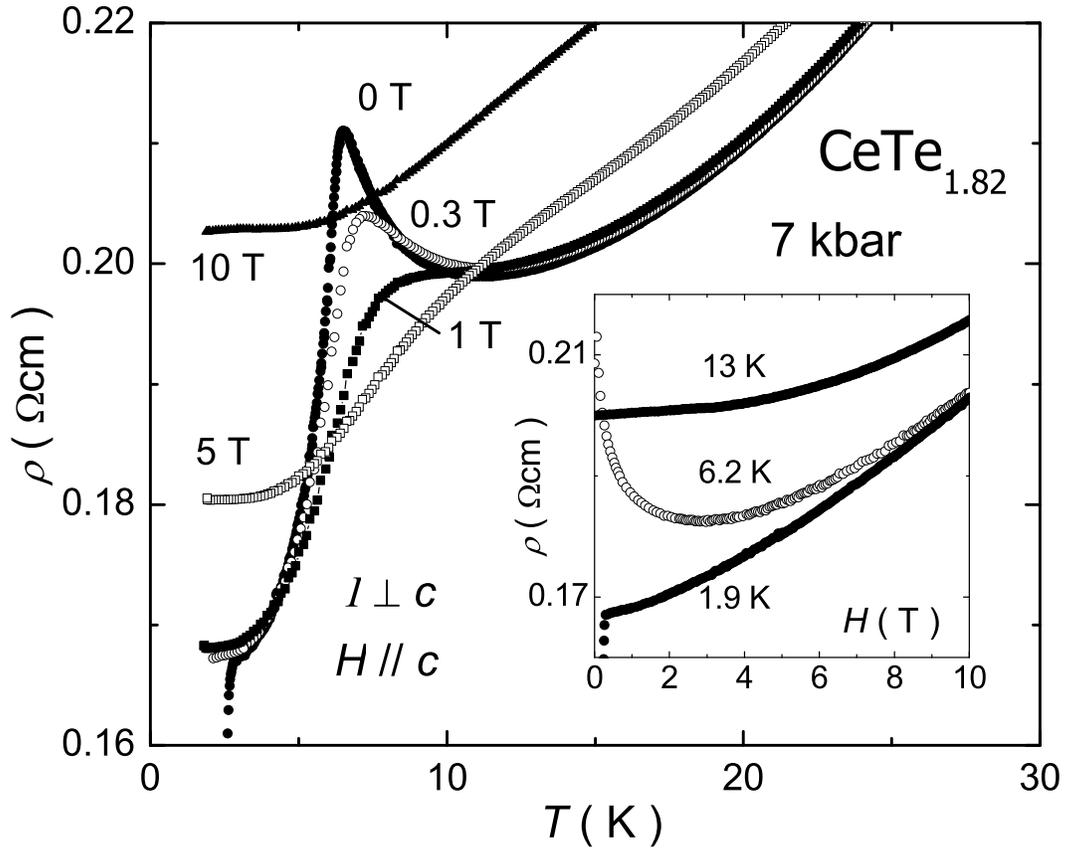}
\caption{In-plane resistivity $\rho_{\perp c}(T)$ for the various
$c$-axis magnetic fields, 0, 0.3, 1, 5, and 10 T. The inset shows
the magnetoresistance (MR) measured at different temperatures,
1.9, 6.2, and 13 K.}
\end{center}
\end{figure}

\begin{figure}
\begin{center}
\includegraphics[width=1\linewidth]{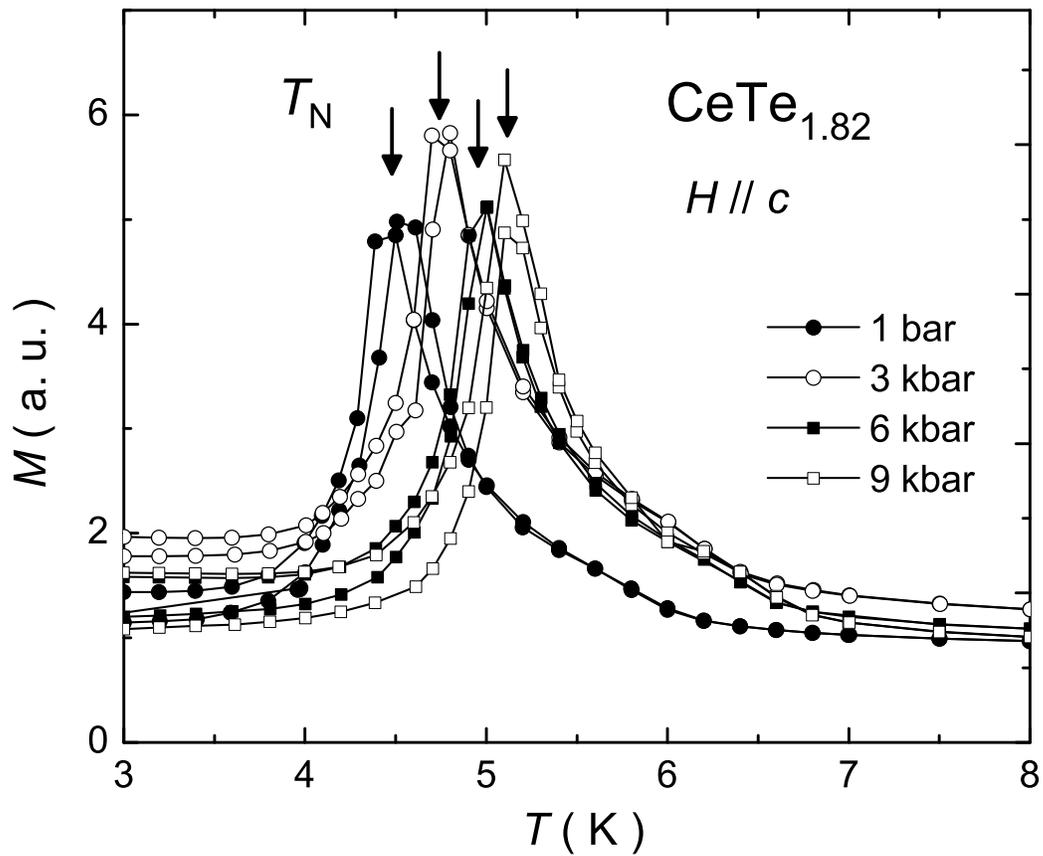}
\caption{Magnetization measured in a field of 50 Oe at various
pressures, 1 bar, 3, 6, and 9 kbar. The arrows indicate the
long-range antiferromagnetic transition temperature, $T_{\rm N}$.}
\end{center}
\end{figure}

\end{document}